\documentclass[prd,aps,showpacs,superscriptaddress,nofootinbib,eqsecnum,amsfonts,amsmath,preprintnumbers,floatfix,floats]{revtex4}

\usepackage{epsfig}
\usepackage{bm}
\usepackage{graphics}
\usepackage{amssymb}
\usepackage[usenames]{color}

\def\bea{\begin{eqnarray}}
\def\eea{\end{eqnarray}}
\def\vt{\vartheta}

\begin{document}

\newcommand{\rhat}{\hat{r}}
\newcommand{\iotahat}{\hat{\iota}}
\newcommand{\phihat}{\hat{\phi}}
\newcommand{\h}{\mathfrak{h}}
\newcommand{\be}{\begin{equation}}
\newcommand{\ee}{\end{equation}}
\newcommand{\ber}{\begin{eqnarray}}
\newcommand{\eer}{\end{eqnarray}}
\newcommand{\fmerg}{f_{\rm merg}}
\newcommand{\fcut}{f_{\rm cut}}
\newcommand{\fring}{f_{\rm ring}}
\newcommand{\cA}{\mathcal{A}}
\newcommand{\ie}{i.e.}
\newcommand{\df}{{\mathrm{d}f}}
\newcommand{\rmi}{\mathrm{i}}
\newcommand{\rmd}{\mathrm{d}}
\newcommand{\rme}{\mathrm{e}}
\newcommand{\dt}{{\mathrm{d}t}}
\newcommand{\pj}{\partial_j}
\newcommand{\pk}{\partial_k}
\newcommand{\psifl}{\Psi(f; {\bm \lambda})}
\newcommand{\hp}{h_+(t)}
\newcommand{\hc}{h_\times(t)}
\newcommand{\Fp}{F_+}
\newcommand{\Fc}{F_\times}
\newcommand{\Ylm}{Y_{\ell m}^{-2}}
\def\no{\nonumber \\ & \quad}
\def\noQ{\nonumber \\}
\newcommand{\mc}{M_c}
\newcommand{\vek}[1]{\boldsymbol{#1}}
\newcommand{\vdag}{(v)^\dagger}
\newcommand{\bvtheta}{{\bm \vartheta}}
\newcommand{\btheta}{{\bm \theta}}
\newcommand{\brho}{{\bm \rho}}
\newcommand{\pa}{\partial_a}
\newcommand{\pb}{\partial_b}
\newcommand{\Psieff}{\Psi_{\rm eff}}
\newcommand{\Aeff}{A_{\rm eff}}
\newcommand{\deff}{d_{\rm eff}}
\newcommand{\corr}{\mathcal{C}}
\newcommand{\bvthat}{\hat{\mbox{\boldmath $\vt$}}}
\newcommand{\bvt}{\mbox{\boldmath $\vt$}}

\newcommand{\comment}[1]{{\textsf{#1}}}
\newcommand{\ajith}[1]{\textcolor{magenta}{\textit{Ajith: #1}}}
\newcommand{\sukanta}[1]{\textcolor{blue}{\textit{Sukanta: #1}}}

\newcommand{\AEIHann}{Max-Planck-Institut f\"ur Gravitationsphysik 
(Albert-Einstein-Institut) and Leibniz Universit\"at Hannover, 
Callinstr.~38, 30167~Hannover, Germany}
\newcommand{\WSU}{Department of Physics \& Astronomy, Washington State University,
1245 Webster, Pullman, WA 99164-2814, U.S.A. \\
}
\newcommand{\LIGOCaltech}{LIGO Laboratory, California Institute of Technology, 
Pasadena, CA 91125, U.S.A.}
\newcommand{\TAPIR}{Theoretical Astrophysics, California Institute of Technology, 
Pasadena, CA 91125, U.S.A.}

\title{A blind hierarchical coherent search for gravitational-wave signals from coalescing compact binaries in a network of interferometric detectors}

\preprint{LIGO-P1000183}

\author{Sukanta Bose}
\email{sukanta@wsu.edu}
\affiliation{\WSU}

\author{Thilina Dayanga}
\email{wdayanga@wsu.edu}
\affiliation{\WSU}

\author{Shaon Ghosh}
\email{shaonghosh@mail.wsu.edu}
\affiliation{\WSU}

\author{Dipongkar Talukder}
\email{talukder_d@wsu.edu}
\affiliation{\WSU}


\begin{abstract}

We describe a hierarchical data analysis pipeline for coherently searching for gravitational wave (GW) signals from non-spinning compact binary coalescences (CBCs) in the data of multiple earth-based detectors. This search assumes no prior information on the sky position of the source or the time of occurrence of its transient signals and, hence, is termed ``blind''. The pipeline computes the coherent network search statistic that is optimal in stationary, Gaussian noise. More importantly, it allows for the computation of a suite of alternative multi-detector coherent search statistics and signal-based discriminators that can improve the performance of CBC searches in real data, which can be both non-stationary and non-Gaussian. Also, unlike the coincident multi-detector search statistics that have been employed so far, the coherent statistics are different in the sense that they check for the consistency of the signal amplitudes and phases in the different detectors with their different orientations and with the 
signal arrival times in them. Since the computation of coherent statistics entails searching in the sky, it is more expensive than that of the coincident statistics that do not require it.
To reduce computational costs, the first stage of the hierarchical pipeline constructs coincidences of triggers from the multiple interferometers, by requiring their proximity in time and component masses. 
The second stage follows up on these coincident triggers by computing the coherent statistics. Here, we compare the performances of this hierarchical pipeline with and without the second (or coherent) stage in Gaussian noise. Whereas introducing hierarchy can be expected to cause some degradation in the detection efficiency compared to that of a single-stage coherent pipeline, nevertheless it improves the computational speed of the search considerably. The two main results of this work are: (1) The performance of the hierarchical coherent pipeline on Gaussian data is shown to be better than the pipeline with just the coincident stage. (2) The three-site network of LIGO detectors, in Hanford and Livingston (USA), and Virgo detector in Cascina (Italy) cannot resolve the polarization of waves arriving from certain parts of the sky. This can cause the three-site coherent statistic at those sky positions to become singular.
Regularized versions of the statistic can avoid that problem, but can be expected to be sub-optimal. 
The aforementioned improvement in the pipeline's performance due to the coherent stage is in spite of this handicap.

\end{abstract}
\maketitle

\section{Introduction}

Signals from binaries of neutron stars (NSs) and black holes (BHs) enjoy the prospect of being the first signals to be detected by gravitational wave (GW) detectors \cite{Thorne:1987af}. They are among the best understood of all GW sources and enough number of them are expected to appear in the data of second generation detectors \cite{:2010cfa}. 
The last several science runs at LIGO \cite{Sigg-LIGOstatus-2008}, GEO600 \cite{Grote-GEOstatus-2008}, and Virgo \cite{VirgoStatus-GWDAW2008} have revealed that searches for signals from these compact binary coalescences (CBCs) benefit from the networking of multiple detectors because of the reduction in the rate of accidentals or false alarms, especially, from non-stationary and non-Gaussian noise artifacts. Further, studies with injection of simulated signals show that the estimation of source parameters, such as sky position and wave polarization, is also helped by networks involving detectors at three or more sites around the globe \cite{Ajith:2009fz,Nissanke:2009kt,Fairhurst:2009tc}. This is important since CBCs may not always emit electromagnetic signals that are strong enough to be observable. 

Searches with no prior information on the sky-position of the source or the time of occurrence of its transient signals are termed ``blind''. This paper describes blind CBC search strategies, which must be contrasted with a targeted search method \cite{Harry10a}. An example of the latter case is a search for GW signal triggered by a short-duration gamma-ray burst (GRB). Short GRBs have been conjectured to be associated with NS-NS or NS-BH coalescences \cite{Eichler:1989ve,Paczynski:1991aq}. Owing to an electromagnetic counterpart, the sky-position of the short GRB and the time of arrival of its gamma-ray signal are known in advance for offline searches. This implies that searches for GW signals from these sources require three less parameters to scan, and are, therefore, computationally less expensive. (In reality, one searches over a several-second window around the arrival time of the gamma-ray signal because it is not clear yet how separated the emission of the gamma-ray burst and the binary-object merger are in time \cite{Abbott:2008zzb,:2010uf}.)
Perhaps more significantly, it reduces the probability of false-alarms and, therefore, increases our detection confidence.

In this paper, we address how one tackles both these issues, namely, of increased computational costs and false-alarm rates, affecting a blind search for signals from CBCs with non-spinning components. To reduce the excess computational cost arising from scanning the arrival time, one introduces hierarchical stages in the search pipeline, whereby, first, the triggers of interest are identified in the detectors individually. This is followed by recognizing triggers that are coincident in multiple detectors and then computing network-based statistics for them that reveal their significance as GW candidates. (These hierarchical steps were introduced in Ref. \cite{Abbott:2007xi} and have been used in multiple CBC searches ever since.) The final stage is used to compute the coherent network statistics for these coincident triggers. To address the second problem of increased false-alarms, especially, from non-stationary noise transients, we introduce signal-based multi-detector discriminators that check for consistency of the signals appearing in individual detectors with a CBC source, after accounting for the different orientations of the detectors and the delays in their times of arrival in them.

Past experiments with multi-detector searches for gravitational-wave signals from compact-binary coalescences (CBCs) have shown that the statistics that are optimal in Gaussian and stationary noise (OGSN) cease to be so in real data, in general \cite{Abbott:2007xi,Abbott:2009tt,Abbott:2009qj}. Instead a function of the chi-square-weighted \cite{Allen:2004gu} matched-filter \cite{Sathyaprakash:1991mt} outputs has been found to deliver a better performance \cite{Abbott:2007xi,Abbott:2009tt}. This function is arrived at empirically by comparing the distribution of the matched-filter and chi-square statistics for simulated CBC signal injections with that of the background. These statistics did not, however, use the phase of the matched-filter output to discriminate signals from noise, which a coherent statistic \cite{Bose:1999pj,Pai:2000zt} is equipped to do. We will call the former {\em coincident} statistics.
Their construction has nevertheless helped inspire techniques for obtaining empirically an effective coherent statistic that performs better in real data than the coherent statistic of Refs. \cite{Bose:1999pj,Pai:2000zt}. It is this statistic and its variants, which can be useful in searching non-spinning CBC signals in real data, that we discuss in detail in this paper.
 
In Sec. \ref{sec:stats} we describe the GW signal in a detector and its relation to signals from the same source in detectors at other locations, and with different orientations. We also revisit the OSGN coherent network search statistic to introduce notation and convention followed in the rest of the paper. We then describe new network statistics that are more robust in detector noise that is non-stationary and non-Gaussian. 
In Sec. \ref{sec:pipeline}, we describe the hierarchical search pipeline used to compute the coherent statistics and other alternative network detection statistics and signal-based discriminators. Section \ref{sec:results} presents the results from running this pipeline in simulated data from the LIGO detectors
at Hanford and Livingston, with 4km-long arm-lengths, and the Virgo detector
in Pisa. 

\section{Multi-detector statistics}
\label{sec:stats}

We begin by describing the statistic that is optimal for coherently searching for non-spinning CBC signals in data from multiple detectors when their noise is Gaussian and stationary. The first part of this section gives an alternative derivation of this statistic, as compared to that available in the literature \cite{Bose:1999pj}. In the process, it introduces notation and convention followed here. It also introduces signal parameters and variables used in the coherent search code available in the LIGO (Scientific Collaboration) Algorithm Library LAL \cite{lal}. We then compare that statistic with the aforementioned empirically-motivated multi-detector coincident statistics, which have been applied in real data. 

\subsection{Signal and noise}
\label{subsec:signalnoise}

Consider a non-spinning coalescing compact binary with component masses $m_{1,2}$, such that its total mass is $M=m_1+m_2$ and its reduced mass is $\mu = m_1m_2/M$.
In the restricted post-Newtonian approximation, the two polarizations determining the GW strain are:
\begin{mathletters}%
\label{eq:strainpols}
\begin{eqnarray} 
h_+(t;r,M,\mu,\iota,\varphi_c,t_{c})&=&{G \mathcal{M}\over c^2 r} \left(\frac{t_c -t}{5G\mathcal{M}/ c^3} \right)^{-1/4} 
{1+\cos^2 \iota \over 2} \>\cos[\varphi(t;t_{c},M,\mu) + \varphi_c]\ \ ,\label{hplus} \\
h_\times(t;r,M,\mu,\iota,\varphi_c,t_{c})&=&{G \mathcal{M}\over c^2 r} \left(\frac{t_c -t}{5G\mathcal{M}/ c^3} \right)^{-1/4}  \cos\iota \>\sin [\varphi(t;t_c,M,\mu) + \varphi_c] \ \ , \label{hcross}
\end{eqnarray}
\end{mathletters}%
which depend on $M$, $\mu$, the luminosity distance to the source $r$, the inclination angle of the source's orbital-momentum vector to the line of sight $\iota$, the time of coalescence of the signal $t_c$, and the coalescence phase of the signal $\varphi_c$. Above, $\varphi(t;t_{c},M,\mu)$ is the orbital phase of the binary \cite{Blanchet:1995ez,Blanchet:1996pi}, $\mathcal{M} = \mu^{3/5}M^{2/5}$ is the chirp mass, $G$ is the gravitational constant and $c$ is the speed of light in vacuum. The GW strain in a detector can then be modeled as,
\be
\label{eq:strain}
h(t) = F_+h_+(t) + F_\times h_\times (t) \,,
\ee
where $F_{+,\times}$ are antenna-pattern functions that quantify the sensitivity of the detector to the sky-position and polarization of the source,
\be
\left(\begin{array}{c} F_+ \\ F_\times \end{array}\right) 
= \left(\begin{array}{cc} \cos 2\psi & \sin 2\psi \\ -\sin 2\psi & \cos 2\psi \end{array}\right) 
\left(\begin{array}{c} u \\ v \end{array}\right) \,,
\ee
with $\psi$ being the wave-polarization angle and $u(\alpha,\delta)$ and $v(\alpha,\delta)$ being detector-orientation dependent functions of the source sky-position angles $(\alpha,\delta)$ \cite{Jaranowski:1996hs,Pai:2000zt}.

Following Ref. \cite{Ajith:2009fz}, let us map the CBC signal parameters
$(r, \psi, \iota, \varphi_c)$, into new parameters, $a^{(k)}$, with 
$k=$1,...,4, such that the strain in any
given detector has a {\em linear} dependence on them:
\be
\label{eq:hOfTDomModeInAs}
h(t) = \sum_{k=1}^4 a^{(k)} {\sf h}_k(t) \,,
\ee
where the ${\sf h}_k(t)$'s are completely independent of those four  
parameters. By comparing the above expression for the GW strain with that defined through Eqs. (\ref{hplus}), (\ref{hcross}), and (\ref{eq:strain}), we find
\bea
{\sf h}_1(t) &\propto&  u(\alpha,\delta)\cos[\varphi(t;M,\mu,\alpha,\delta,t_c)] \,, \noQ
{\sf h}_2(t) &\propto&  v(\alpha,\delta)\cos[\varphi(t;M,\mu,\alpha,\delta,t_c)] \,, \noQ
{\sf h}_3(t) &\propto&  u(\alpha,\delta)\sin[\varphi(t;M,\mu,\alpha,\delta,t_c)] \,, \noQ
{\sf h}_4(t) &\propto&  v(\alpha,\delta)\sin[\varphi(t;M,\mu,\alpha,\delta,t_c)] \,,
\eea
where the proportionality factor is $[G \mathcal{M}/c^2] [ (t_c -t)/(5G\mathcal{M}/ c^3)]^{-1/4}$. 
This method of resolving the GW strain signal in a basis of four time-varying functions was first found in Ref. \cite{Jaranowski:1998qm} for pulsar signals.

The new parameters, $a^{(k)}$, with the index $k$ taking four values, are defined in terms of $(r, \psi, \iota, \varphi_c)$ as,
\bea
\label{eq:aDef}
a^{(1)} &=& \frac{1}{r}\left(\cos2\psi \cos\varphi_c\frac{1+\cos^2\iota}{2}-\sin2\psi \sin\varphi_c\cos\iota\right) \,, \noQ
a^{(2)} &=& \frac{1}{r}\left(\sin2\psi \cos\varphi_c\frac{1+\cos^2\iota}{2}+\cos2\psi \sin\varphi_c\cos\iota\right) \,, \noQ
a^{(3)} &=& -\frac{1}{r}\left(\cos2\psi \sin\varphi_c\frac{1+\cos^2\iota}{2}+\sin2\psi \cos\varphi_c\cos\iota\right) \,, \noQ
a^{(4)} &=& -\frac{1}{r}\left(\sin2\psi \sin\varphi_c\frac{1+\cos^2\iota}{2}-\cos2\psi \cos\varphi_c\cos\iota\right) \,.
\eea
These constitute an alternative set of parameters that define the likelihood ratio. We used parenthetic indices above to avoid confusing them with numerical exponents.

\subsection{The network detection statistic}
\label{subsec:netstat}

Let the inner-product of two temporal functions $p(t)$ and $q(t)$ be defined as
\be\label{innerprod}
\langle p ,\>q \rangle_{(I)} = 4 \Re \int_{0}^{\infty} \! \df\>
{\tilde{p}^* (f) \,\tilde{q}(f) \over S_{(I)}(f)} \ \ , 
\ee
where $\tilde{p}(f)$ and $\tilde{q}(f)$ are the Fourier
transforms of $p(t)$ and $q(t)$, respectively, and  $S_{(I)}(f)$ is the
one-sided noise PSD of the $I$th detector \cite{Helstrom}, with $I={1,...,M}$ for a network of $M$ detectors. The angular brackets denoting the inner-product are subscripted with the detector index since that product depends on the noise PSD of the detector. Assuming that detector noise $n^I(t)$ is additive, the strain in a detector in the presence of a CBC signal is 
\be
x^I(t) = n^I(t) + h^I(t) \,,
\ee
where 
$h^I(t)$ is 
given by Eq. (\ref{eq:strain}), but now with the antenna-pattern functions superscripted with the detector index. (The polarization components $h_{+,\times} (t)$ also depend on $I$ through the coalescence time, as explained below.)
Moreover, if the noise is zero-mean Gaussian and stationary, the log-likelihood ratio (LLR) is \cite{Helstrom}
\be
\label{LLRI}
\log \Lambda_I =\langle x^I ,\> h^I\rangle_{(I)} 
-\frac{1}{2}\langle h^I ,\> h^I\rangle_{(I)}  \,,
\ee
which can serve as a statistic for detecting signals in a single detector.

To explore the properties of the LLR, it will be useful to define the (complex) unit-norm template $S^I(t)$ associated with the circular-polarization component of a GW, namely, $h_+(t)+ih_\times(t)$. It can be shown \cite{Pai:2000zt} that
\be
S^I(t) = g_{(I)}^{-1}\left[\xi^I\left( t_c - t \right)\right]^{-1/4}~e^{i\varphi\left( t \right)}\,,
\ee
where $g_{(I)}$ (with units of $\sqrt{\rm Hz}$) is a normalization factor, such that $\langle S^I, S^I \rangle = 1$, and
\be
\xi^I = \frac{5}{256f_s^I} \left[ \frac{G\mathcal{M} f_s^I}{c^3}\right]^{-5/3}
\ee
is the time spent by the signal in the detector band, in the Newtonian approximation. Above, $f_s^I$ is the seismic cut-off frequency of the $I$th detector below which it has little sensitivity for GW signals. The single detector matched-filter output against $S^I(t)$ can then be defined as
\be
\label{eq:rho}
C^I = \langle S^I, x^I \rangle \equiv \left(c_+^I + i c_-^I \right) = \rho^I e^{i\phi^I} \,,
\ee
where $c_\pm^I$, $\rho^I$ and $\phi^I$ are all real; $\rho^I = |C^I|$ is often termed as the signal-to-noise ratio (SNR) in the $I$th detector. Since the detector strain due to a GW signal is expected to be tiny, one has $g_{(I)} \gg 1$. Therefore, for computational efficiency, we define a new factor that is closer to unity,
\be
\sigma_{(I)} \equiv \left(\frac{G\mathcal{M}/c^2}{\rm 1~Mpc}\right) \left(\frac{5G\mathcal{M}\xi}{c^3}\right)^{1/4}~g_{(I)} \,,
\ee
with $\xi$ computed for a reference detector selected from one of those in the network. This is convenient since, as explained below, the detection statistics and the  parameters $\{\psi,\iota,\varphi_c\}$ are all independent of the above parenthetic scale factors; only the source distance depends on them, and is computed after accounting for them.

Using the strain expression in Eq. (\ref{eq:hOfTDomModeInAs}), the LLR for a network of multiple detectors can be recast in terms of $a^{(k)}$, provided one knows how the strain from the same CBC signal varies from one detector to the other. This was explained in Refs. \cite{Bose:1999pj,Pai:2000zt}. Here, it suffices to note that this dependence arises owing to: (a) The spatial separation of the detectors, which can cause relative delays in the arrival of the signal. These delays are determined by the source's sky-position and can be accounted for in Eqs. (\ref{hplus}) and (\ref{hcross}) by adding those delays to $t_c$. (b) The different orientations of the detectors, which change $u$ and $v$. 
Assuming that the noise in the different detectors are statistically independent, the joint log-likelihood ratio for a network of $M$ detectors is
\bea
\log\left({}^{(M)}\Lambda\right) &=& \sum_{I=1}^M \log \Lambda_I \nonumber \\
&=& N_ka^{(k)} - \frac{1}{2}M_{ij}a^{(i)}a^{(j)} \,, \label{eq:netLLR}
\eea
where, in the last expression, the sum over detectors has been absorbed in $N_k$ and $M_{ij}$, as defined below:
\be
\left(\begin{array}{c} N_1 \\N_2 \\N_3 \\N_4 \end{array}\right)
= \chi\left(\begin{array}{c} \sum_{I=1}^M \sigma_{(I)} u_I c_+^I \\
\sum_{I=1}^M \sigma_{(I)} v_I c_+^I \\
\sum_{I=1}^M \sigma_{(I)} u_I c_-^I \\
\sum_{I=1}^M \sigma_{(I)} v_I c_-^I \end{array} \right)
= \chi\left(\begin{array}{c} 
{\bf u}_\sigma \cdot {\bf c}_+ \\
{\bf v}_\sigma \cdot {\bf c}_+ \\
{\bf u}_\sigma \cdot {\bf c}_- \\
{\bf v}_\sigma \cdot {\bf c}_-
\end{array}
\right)\,.
\ee
Above, 
${\bf u}_\sigma$ and ${\bf v}_\sigma$ are network vectors with components $\sigma_{(I)} u_I$ and $\sigma_{(I)} v_I$, respectively, ${\bf c}_\pm$ are network vectors with components $c_\pm^I$, and 
\be
\chi \equiv \pi^{2/3} \left[\frac{GM_{\odot}/c^2}{1 {\rm Mpc}}\right]^{3/4}~{\rm Mpc} 
\ee
is a normalization factor with dimensions of length. Also,
\be
{\bf M} = \left(\begin{array}{cccc} A & B & 0 & 0 \\
B & C & 0 & 0 \\
0 & 0 & A & B \\
0 & 0 & B & C \end{array}\right)
\ee
with
\be
\left(\begin{array}{c} A \\ B \\ C \end{array}\right)
= \chi^2 \left(\begin{array}{c}
\|{\bf u}_\sigma\|^2 \\
{\bf u}_\sigma \cdot {\bf v}_\sigma \\
\|{\bf v}_\sigma\|^2
\end{array}\right)\,,
\ee
which define the network template-norm, namely, twice the second term on the right-hand side of Eq. (\ref{eq:netLLR}); the first term there can be interpreted as the matched-filter output of the network data-vector, ${\bf x} \equiv \{x^1, x^2,...,x^M\}$ \cite{Pai:2000zt}. 

Maximizing $2\log~{}^{(M)}\Lambda$ with respect to ${\bf a} = \{a^{(1)},a^{(2)},a^{(3)},a^{(4)}\}$ yields 
\be
\label{eq:netMLR}
2\log~{}^{(M)}\Lambda\Big|_{\bar{\bf a}} ={\bf N}^T\cdot {\bf M}^{-1}\cdot {\bf N} \,,
\ee
which 
is still a function of $\{M,\mu,\alpha,\delta,t_c\}$. (Note that the above statistic is independent of $\chi$.) The concomitant maximum likelihood {\em estimates} (MLEs) of the complementary set of four parameters are denoted with an overline:
\be
\label{eq:MLEGen}
\bar{\bf a} =  {\bf M}^{-1}\cdot {\bf N}\,.
\ee
These estimates are also functions of $\{M,\mu,\alpha,\delta,t_c\}$, and are determined by the data through $c_{\pm}^I$ as follows:
\be
\label{eq:aMLE}
\left(\begin{array}{c} \bar{a}^{(1)} \\\bar{a}^{(2)} \\\bar{a}^{(3)} \\\bar{a}^{(4)} \end{array}\right)
= \frac{\chi}{\Delta}\left(\begin{array}{c} 
\|{\bf v}_\sigma\|^2 ~\left( {\bf u}_\sigma \cdot {\bf c}_+ \right)
  -  \left({\bf u}_\sigma \cdot {\bf v}_\sigma\right)  ~\left( {\bf v}_\sigma \cdot {\bf c}_+ \right) \\
- \left({\bf u}_\sigma \cdot {\bf v}_\sigma\right) ~ \left( {\bf u}_\sigma \cdot {\bf c}_+ \right)  + \|{\bf u}_\sigma\|^2 ~\left( {\bf v}_\sigma \cdot {\bf c}_+ \right) \\
\|{\bf v}_\sigma\|^2 ~\left( {\bf u}_\sigma \cdot {\bf c}_- \right)
  -  \left({\bf u}_\sigma \cdot {\bf v}_\sigma\right)  ~\left( {\bf v}_\sigma \cdot {\bf c}_- \right) \\
- \left({\bf u}_\sigma \cdot {\bf v}_\sigma\right) ~ \left( {\bf u}_\sigma \cdot {\bf c}_- \right)  + \|{\bf u}_\sigma\|^2 ~\left( {\bf v}_\sigma \cdot {\bf c}_- \right)
\end{array}\right)\,,
\ee
where $\Delta \equiv AC - B^2$. The MLE of a parameter will be denoted by placing an overline on its symbol.

It is important to note that the maximization in Eq. (\ref{eq:netMLR}) assumes that the network matrix ${\bf M}$ is invertible. This is not true, in general. Indeed, ${\bf M}$ is singular when ${\bf u}_\sigma$ is aligned with ${\bf v}_\sigma$. These two vectors are determined by how the interferometers in the network are oriented with respect to the wave propagation vector, but are not affected by the polarization angle $\psi$. In addition to this singularity, ${\bf M}$ can be rank deficient, thus, making the problem of inverting it ill-posed \cite{Rakhmanov:2006qm}. Physically, this implies that the network does not have enough linearly independent basis detectors to be able to resolve the source parameters ${\bf a}$. Note that these maladies of ${\bf M}$ are dependent on the sky-position angles. This means that a network that is able to resolve the signal parameters for certain source sky-positions may not be able to do so for others. These problems can be tackled by regularizing ${\bf M}$ in a variety of ways that have been explored in the context of searches of transient signals from unmodeled sources, also called ``burst'' searches \cite{Rakhmanov:2006qm,Klimenko:2005xv,Mohanty:2006ha}. These methods obviate the rank-deficiency problem at the cost of making the search statistic sub-optimal. Thus, any deficiencies arising from potential singularities in ${\bf M}$ or its regularization method adopted by a search pipeline will affect its performance. Since ${\bf M}$ is independent of the detector strain data, such effects will arise in searches in simulated Gaussian data sets as well, such as the ones studied here. Since our results below are devoid of these maladies, we are confident that they will not arise in real data searches as well.

The maximum-likelihood estimates for the four physical parameters $(r, \psi, \iota, \varphi_c)$ can now be expressed in terms of the above estimates by inverting Eq. (\ref{eq:aDef}) and replacing ${\bf a}$ with $\bar{\bf a}$. Specifically, for the luminosity distance we get:
\be
\bar{r} = \frac{\sqrt{1 + 6\cos^2\bar{\iota} + \cos^4\bar{\iota}}}{2\|\bar{\bf a}\|}\,,
\ee
where $\|\bar{\bf a}\| \equiv \sqrt{\sum_{i=1}^4\left(\bar{a}^{(i)}\right)^2}$ is the norm of the four-parameter vector MLE, and $\bar{\iota}$ is defined below along with the other MLEs. Since those angular parameter estimates should not depend on an overall scaling of $\bar{\bf a}$,
it helps to define the dimensionless unit-norm components
$\hat{\bar{a}}^{(k)} \equiv \bar{a}^{(k)} / \|\bar{\bf a}\|$.
In terms of the $\hat{\bar{a}}^{(k)}$, the maximum-likelihood estimates for the three angular parameters are,
\bea
\label{eq:angMLEs}
\bar{\psi} &=& \frac{1}{4}\sin^{-1}\left( \frac{2\left(\hat{\bar{a}}^{(1)}\hat{\bar{a}}^{(2)}+\hat{\bar{a}}^{(3)}\hat{\bar{a}}^{(4)}\right)}{\sqrt{1-\zeta^2}}\right)\,, \noQ
\bar{\phi}_c &=& -\frac{1}{2}\sin^{-1}\left( \frac{2\left(\hat{\bar{a}}^{(1)}\hat{\bar{a}}^{(3)}+\hat{\bar{a}}^{(2)}\hat{\bar{a}}^{(4)}\right)}{\sqrt{1-\zeta^2}}\right)\,, \noQ
\bar{\iota} &=& \cos^{-1}\left(\frac{1-\sqrt{1-\kappa^2}}{\kappa}\right) \,,
\eea
where $\zeta \equiv 2\left(\hat{\bar{a}}^{(1)}\hat{\bar{a}}^{(4)}-\hat{\bar{a}}^{(2)}\hat{\bar{a}}^{(3)}\right)$ and
\be
\kappa = \frac{\zeta}{1+\sqrt{1-\zeta^2}}\,.
\ee
Note that the expression for $\bar{\psi}$ goes over to that of $\bar{\phi}_c$ under the transformation $\bar{\psi} \longrightarrow (-\bar{\phi}_c)/2$ and $\hat{\bar{a}}^{(2)} \leftrightarrow \hat{\bar{a}}^{(3)}$. This relation arises from a similar symmetry exhibited by the $a^{(k)}$ defined in Eq. (\ref{eq:aDef}).
Expressions for the CBC MLEs and the coherent statistic were first obtained in Refs. \cite{Bose:1999pj,Pai:2000zt}. Above, we reexpress them in terms of the four parameters $a^{(k)}$ since the search code in LAL uses them \cite{lal}. 

Substituting for ${\bf M}$ and ${\bf N}$, the MLR can be expanded as,
\be
\label{eq:cohsnrexp}
2\log \Lambda\Big|_{\bar{\bf a}} =
\left( {\bf w}_+\cdot {\bf c_+} \right)^2 + \left( {\bf w}_-\cdot {\bf c_+} \right)^2 + \left( {\bf w}_+\cdot {\bf c_-} \right)^2 + \left( {\bf w}_-\cdot {\bf c_-} \right)^2\,,
\ee
where ${\bf w}_\pm$ are network vectors with components $w_\pm^I$,
\be
\left(\begin{array}{c} w_{I+}  \\ w_{I-}  \end{array}\right)
= \left(\begin{array}{cc} O_{11} & O_{12} \\
O_{21} & O_{22} \end{array}\right) 
\left(\begin{array}{c}
\sigma_{(I)} u_I \\ \sigma_{(I)} v_I \end{array}\right) \,,
\ee
and
\be
\left(\begin{array}{cc} O_{11} & O_{12} \\
O_{21} & O_{22} \end{array}\right) 
= 
\frac{1}{\sqrt{2\Delta}}
\left(\begin{array}{cc}
{\sqrt{C+A+D}}/{G_1} &
~~~{\sqrt{C+A+D}(C-A-D)}/(2BG_1)  \\
{\sqrt{C+A-D}}/{G_2} &
~~~{\sqrt{C+A-D}(C-A+D)}/(2BG_2) 
\end{array}\right) \,,
\ee
with $D \equiv \sqrt{ (A-C)^2 + 4B^2}$ and
$G_{1,2} \equiv \sqrt{\left( C- A \mp D\right)^2 + 4B^2}~/(2B)$.
The above matrix diagonalizes ${\bf M}$ and, in so doing, identifies the dominant polarization basis, first identified in \cite{Bose:1999pj} and named as such in \cite{Klimenko:2005xv}. 

The coherent search statistic is just $2\log \Lambda\Big|_{\bar{\bf a}}$ maximized over $\{M,\mu,\alpha,\delta,t_c\}$, namely,
\be
\label{eq:cohstat}
\rho_{\rm coh}^2 = 2\log \Lambda\Big|_{\bar{\bvtheta}}\,,
\ee
where $\bvtheta = \{a^{(1)},a^{(2)},a^{(3)},a^{(4)},M,\mu,\alpha,\delta,t_c\}$ is a set of nine parameters for the non-spinning CBC signal. The last five parameters are searched for numerically, by using a grid for the masses and the sky-position and by using the fast Fourier transform \cite{NRecipes} to search for the coalescence time. $\bar{\bvtheta}$ denotes the MLE values of these parameters. Searching over $(\alpha,\delta)$ requires the flexibility to delay $c^I_\pm$ relative to $c^J_\pm$ by an interval that can be anywhere between zero and the light-travel-time between the locations of the $I$th and $J$th detectors or the negative of it. This is why we construct small snippets of $C^I(t)$
called {\em C-data} around the end-time of every trigger that is found to be coincident in multiple detectors in a network. The statistic defined above will be termed as the coherent network SNR and is the detection statistic optimal in stationary, Gaussian noise \cite{Pai:2000zt}.

On the other hand, the {\em combined} signal-to-noise ratio, which was used as a detection statistic in the past and is used here in Fig. \ref{fig:combinedVsCoherentSNR} for comparison, is defined as 
\be
\rho_{\rm comb}^2 = \sum_{I=1}^M \left( \rho^I \right)^2 = \|{\brho}\|^2\,,
\ee
which is devoid of two significant pieces of information present in the coherent search statistic in Eq. (\ref{eq:cohstat}). The first piece of information is in the form of the $w_{I\pm}$ factors, which assign more weight to the matched-filter output of the detector that is more sensitive to a given sky-position and has a lower noise PSD (or bigger $\sigma_{(I)}$). The second piece of information is in the form of the cross-detector terms that check for the consistency of the phases $\phi^I$ with those expected of a real signal.

\subsection{Alternative statistics}
\label{sec:altstats}

The last several science runs at LIGO, GEO600, TAMA, and Virgo
have shown time and again that real detector data is both non-stationary and non-Gaussian. Consequently, neither the single-detector matched-filter-based SNR nor the coherent network SNR are optimal in that data. It is also known that empirically constructed search statistics perform better there. These alternative search statistics are based on signal discriminators such as the chi-square \cite{Allen:2004gu} and rho-square tests \cite{Rodriguez07}, and their performances are compared against the statistics that are optimal in Gaussian and stationary noise. These performances are evaluated in terms of their receiver-operating characteristics, which in turn are constructed from detection efficiencies of simulated signals injected into network data and from the background rates obtained through multiple time-slide experiments.

The statistic that performs better in single-ifo searches is the matched-filtered output weighted by a function of the $\chi^2$ (or chi-square) statistic \cite{Abbott:2007xi},
\be
\rho_{\rm eff} \equiv \rho~\left[\frac{\chi^2}{\left(2p_{\chi^2}-2 \right)}\left(1 + \frac{\rho^2}{\rho_o^2}\right)\right]^{-1/4} \,,
\ee
where $\rho_0$ and $p_{\chi^2}$ are empirical parameters that are deduced by examining the performance of $\rho_{\rm eff}$ in real data. In the latest {\em low-mass} LIGO search, with $2M_\odot<M<35M_\odot$, they were chosen to be 250 and 16, respectively \cite{Abbott:2009qj}. For the high-mass ($25M_\odot<M<100M_\odot$) search studied below, these choices are 50 and 10, respectively. Here, $p_{\chi^2}$ is the number of degrees of freedom of the chi-square statistic, and $\rho_o$ is chosen so that for small $\rho$ and average chi-square values,  $\rho_{\rm eff} \approx \rho$. A large chi-square value indicates that the disagreement between the PSDs of the search template and the putative signal (or noise artifact) in the data is large, and imparts a greater penalty on $\rho_{\rm eff}$ by reducing its value relative to $\rho$.

The network equivalent of the effective SNR is
\be
{}^{(M)}\rho_{\rm eff} = \sqrt{\sum_{I=1}^M \left(\rho_{\rm eff}^I\right)^2}
\ee
and is defined this way simply because it works in real data in discriminating signal injections from background.
A coherent statistic that can perform better in real data than its OGSN kin is constructed straightforwardly by replacing $c_\pm^I$ with 
\be
c_{\pm {\rm eff}}^I \equiv c_\pm^I~\left[\frac{\chi_I^2}{\left(2p_{\chi_I^2}-2 \right)}\left(1 + \frac{\left(\rho^I\right)^2}{\left(\rho_o^I\right)^2}\right)\right]^{-1/4} \,,
\ee
in Eq. (\ref{eq:cohsnrexp}).
Since the $\rho^I$ and $\chi_I^2$ statistics are computed in the CBC search pipeline when the data from the individual detectors are filtered, their values are available to the coherent stage for computing the {\em chi-square-weighted} coherent statistic defined above at little additional computational cost.

Scrutinizing expression (\ref{eq:cohsnrexp}) of $\rho_{\rm coh}$, one finds that it can be decomposed into two parts. The first part is
\be\label{eq:autocoh}
\rho_{\rm auto-coh}^2 =  \sum_{I=1}^M \left( w_{I+}^2 + w_{I-}^2 \right) \left| C^I \right|^2
\ee
and is a sum of auto-correlation terms in each detector. This part of the coherent statistic is less discriminatory between signal and noise triggers.
The second part,
\be\label{eq:crosscoh}
\rho_{\rm cross-coh}^2 =  \sum_{I=1}^M \sum_{\substack{J=1\\\left(J\neq I\right)}}^M \left( w_{I+} w_{J+} + w_{I-} w_{J-} \right)
 \left[  c_{+}^I c_{+}^J + c_{-}^I c_{-}^J \right]\,,
\ee
by contrast, is a sum of cross-correlation terms across pairs of detectors, or baselines, and is critical in checking for phase consistency among signals appearing in the detectors from a GW source. Once again, both of the above statistics can be made more robust against noise glitches by replacing $c_\pm^I$ with $c_{\pm~{\rm eff}}^I$ to obtain their chi-square-weighted counterparts.

Another statistic that is helpful in discriminating signals from noise glitches in multi-detector data is the null-stream \cite{Guersel:1989th}. If $\tilde{C}^I(f)$ is the Fourier transform of $C^I(t)$, then one can show that for GW signals in the data, the mean of
\be
\label{eq:trueNS}
Y \equiv \sum_{I=1}^M K_I\sigma_{\rm inv}^{(I)}S_{h(I)}(f) \tilde{C}^I(f)
\ee
is zero. Above, $K_I = \epsilon_{IJK}F_+^JF_\times^K$, with $\epsilon_{IJK}$ being the Levi-Civita symbol, and $\sigma_{\rm inv}^{(I)} \equiv (\sigma_{(I)})^{-1}$. For non-stationary artifacts, however, this need not be true, thereby, motivating the following discriminator:
\be
\eta = \frac{\langle |Y| \rangle}{\sqrt{{\rm Var}\left( |Y| \right)}}\,,
\ee
where $\langle x \rangle$ and ${\rm Var}(x)$ denote the statistical average and variance of $x$, respectively.
The above construct is called the null-stream statistic. Just like the coherent
SNR, it can be decomposed into two parts as well, comprising auto-correlation and cross-correlation terms, respectively. The former
is akin to the incoherent energy defined in Ref. \cite{Chatterji:2006nh}
for burst searches and will be denoted as $\eta_{\rm auto}$. 
For GW signals one expects $\eta_{\rm auto}$ to be large while $\eta$ itself is small. On the other hand, for noise artifacts, $\eta$ is expected to be large, on the average, even when $\eta_{\rm auto}$ itself is not very strong. This analysis argues for a new statistic, namely,
\be
\label{eq:ratiostat}
R = \eta_{\rm auto} / \eta \,,
\ee
which we call the ratio-statistic. This is a yet another contender for an alternative statistic that can prove useful in real data searches \cite{Ghosh}.

\section{The coherent hierarchical inspiral analysis pipeline}
\label{sec:pipeline}

The coherent hierarchical inspiral analysis (CHIA) pipeline mainly comprises two stages, namely, the coincident and coherent stages, respectively. Both involve multiple steps. The coincident stage has been discussed in the past in Refs. \cite{Abbott:2007xi,Abbott:2009tt}
and is described here briefly for completeness. It includes the following steps: (a) Compute noise PSDs and generate template-banks of the two component masses for each detector in the network. The noise PSDs vary from one detector to another, and in time. 
A template bank is constructed for every 2048s chunk of data from every detector \cite{Brown04}. (b) Use the template bank for each detector to filter the data from that detector and output the parameters of triggers crossing the chosen SNR threshold. For the injection studies, simulated software-injections are added in software to the data in this step, before the data are match-filtered. (c) Parameters of the triggers from the participating detectors are then compared to identify coincidences \cite{Robinson:2008un}.
Before these coincident triggers are considered as detection candidates, in real data one usually applies data-quality vetoes. For our study in simulated data, we forego this stage of the pipeline and, instead, apply the coherent stage directly to the triple-coincident triggers. For the computation of the coherent and null-stream statistics the C-data time-series, which include both the amplitude and the phase time-series of the matched-filter outputs, are required. These time-series are computed in the coherent stage and not upstream in the pipeline since it is computationally less expensive to identify coincidences and construct the C-data only for them.
 
The coherent stage in the CBC search pipeline is constituted of 4 steps. In the first step, a ``coherent bank'' of templates is constructed from the parameters of the coincident triggers. Triggers in different detectors that are coincident and arise from the same GW source can have different mass pairs owing to the possibility that the noise PSDs of the detectors they arise in are somewhat different and because of the random nature of noise. 
For every coincident trigger we construct a network template with a single mass-pair, namely, the one corresponding to the loudest SNR among all the detectors, to search coherently around the end-time of that putative signal. This mass-pair will be termed as the {\em max}-SNR pair and the corresponding detector the {\em max}-SNR detector. For example, consider a triple-coincident trigger with $\{\rho,m_1/M_\odot,m_2/M_\odot\}$ = $\{10.0,1.43,1.39\}$,  $\{10.9,1.40,1.36\}$, and $\{8.9,1.51,1.32\}$ in the first, second, and third interferometric detector (or IFO), respectively. Then the {\em max}-SNR detector is IFO-2 and the {\em max}-SNR mass-pair is $\{m_1/M_\odot,m_2/M_\odot\}$ = $\{1.40,1.36\}$, which is the template included in the coherent bank to represent this coincident trigger in the coherent stage.

While this mass pair will not necessarily give the loudest SNR in the two other detectors, it has been found to yield a better performance for the coherent-statistic and null-stream than when they are computed using the original and, often, non-identical mass pairs in the different detectors. (Note, however, that simulated software injections in real data must used to determine empirically if the detection efficiency is helped by using the same mass pair across all detectors in any given science run.) Also, since error-covariances are known to exist between the mass parameters and the trigger end-time, we search at and {\em around} the end-times of the single-detector triggers that constitute a given network trigger. 

\begin{figure*}[tb]
\centering
\includegraphics[width=8.5cm]{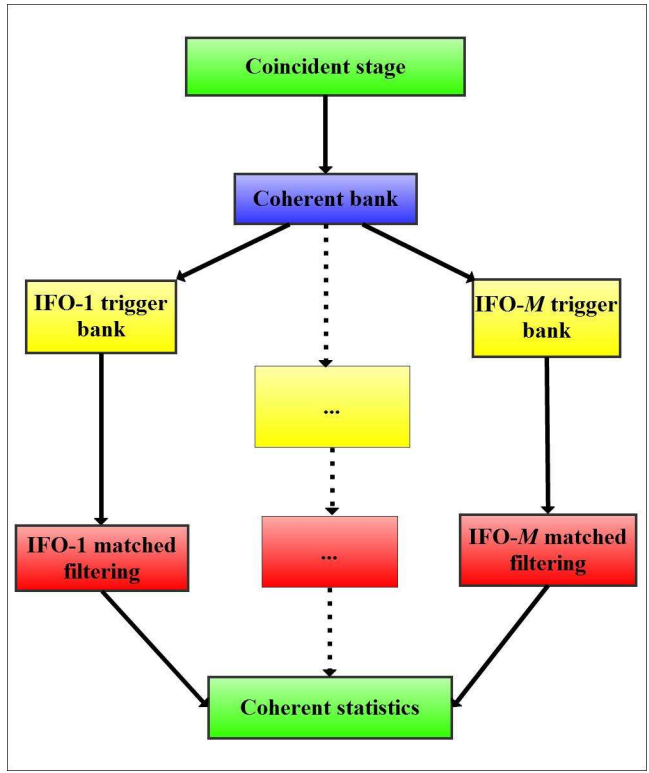}
\caption{
A schematic diagram of the coherent stage in the compact binary coalescence search pipeline.
}
\label{fig:pipeline}
\end{figure*}

The second step in the coherent stage is the construction of {\em trigger-banks}, whereby the coherent-bank template for every coincident trigger is copied as a single-detector template. (See Fig. \ref{fig:pipeline}.)
In the subsequent step, the single-detector templates are used to filter the data from the individual IFOs. This step outputs the 
time-series of C-data around the trigger end-times in that detector. Additionally, this step computes the template normalization factor and chi-square for the {\em max}-SNR mass-pair across all detectors per coincident trigger. Note that the values of these constructs are not available earlier in the pipeline for the triggers in the detectors complementary to the {\em max}-SNR detector since, in general, the mass-pairs would be somewhat different in the preceding coincident stage of the search pipeline. In summary, this step outputs a C-data time-series and the corresponding signal parameters, such as the template normalization factor, for every trigger listed in the {\em coherent-bank} output file.

The final step of the coherent stage is the {\em coherent-statistics} step, which matches the parameters of each triple-coincident trigger to the C-data time-series output by the matched-filtering step and uses them and the corresponding template-norms, chi-square values for the respective detectors to compute a variety of multi-detector statistics, such as the coherent SNR, null-stream, the chi-square-weighted coherent SNR, and other alternative statistics.

\section{Results}
\label{sec:results}

\begin{figure*}[tb]
\centering
\includegraphics[width=8.5cm]{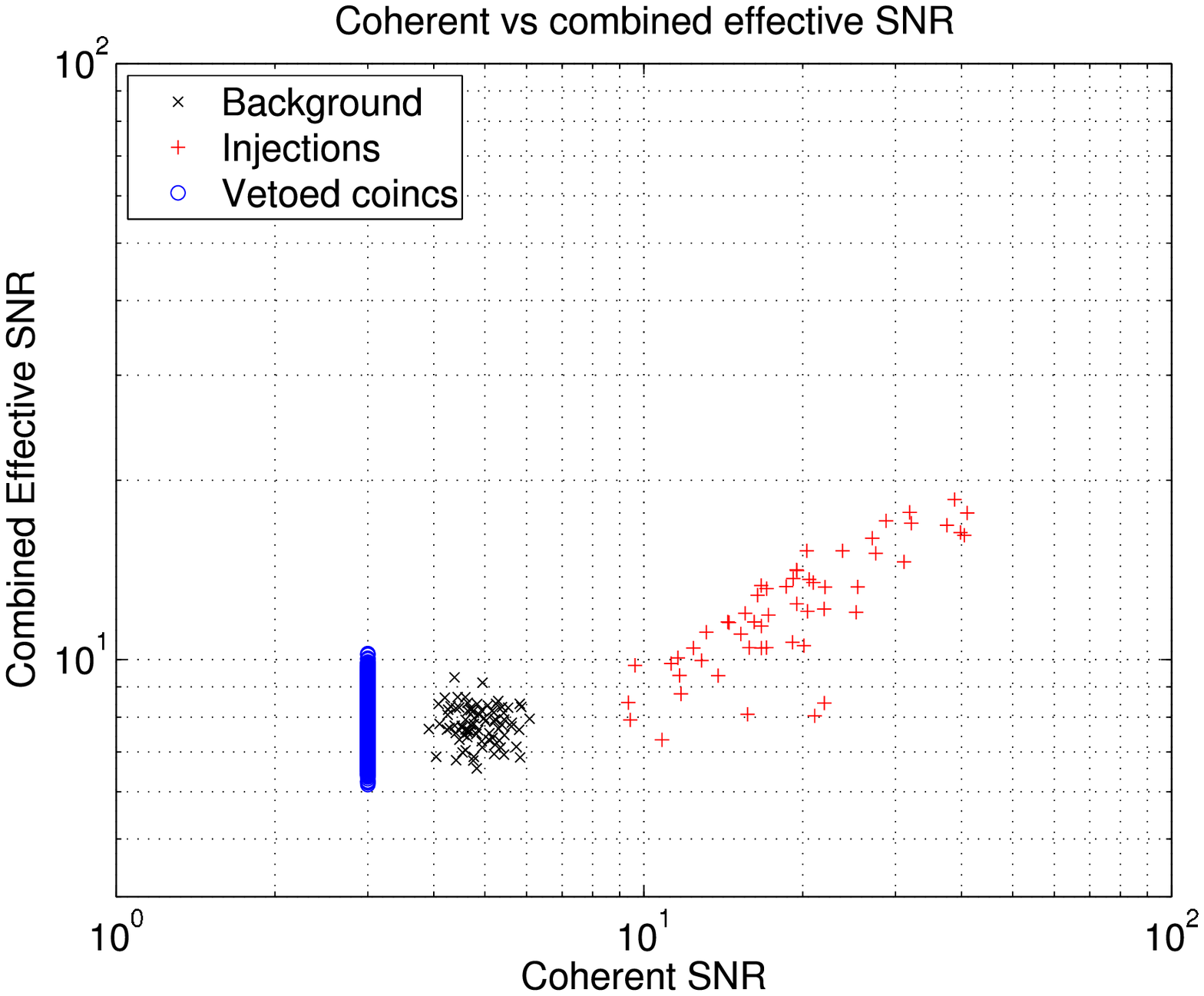}
\includegraphics[width=8.5cm]{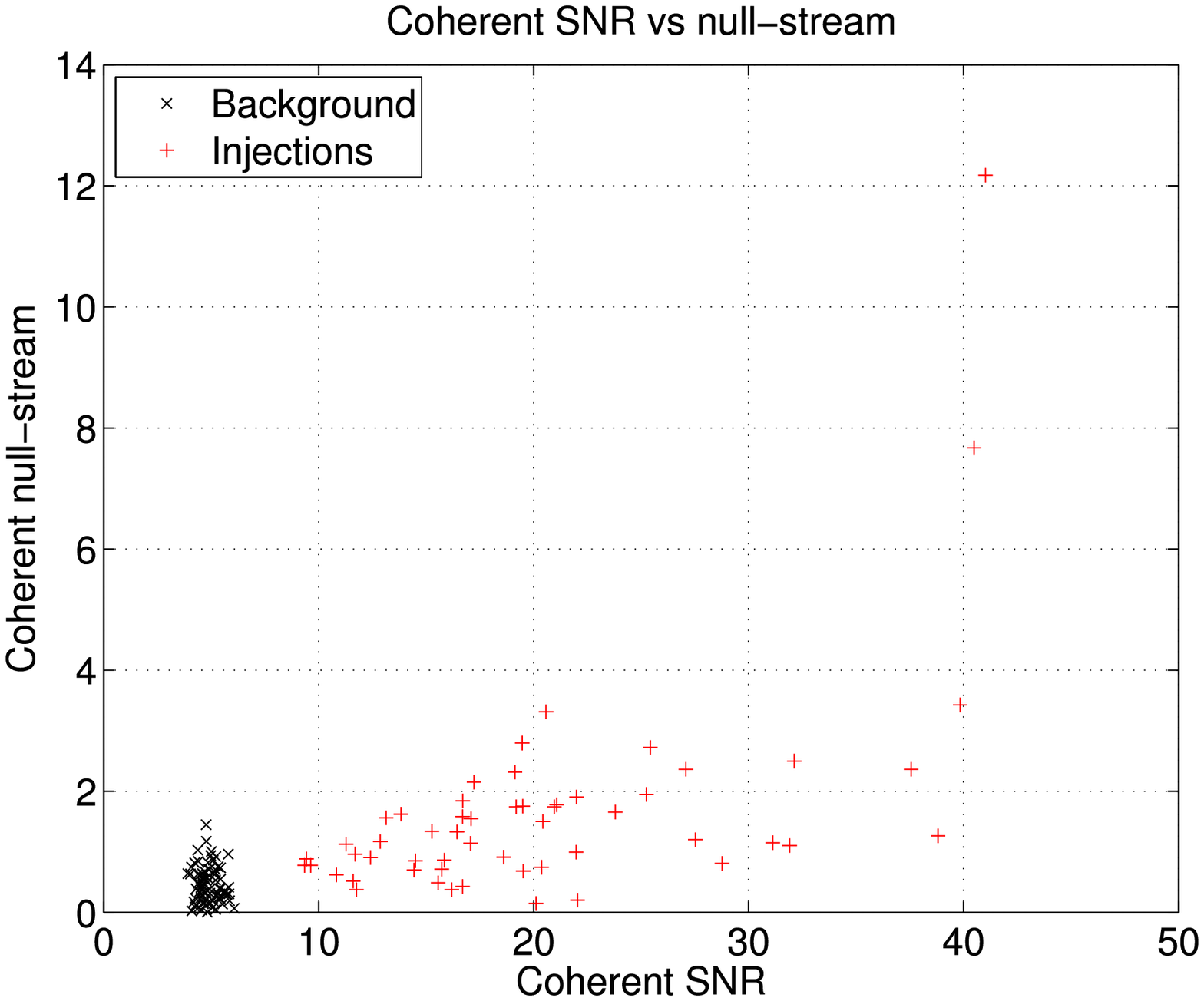}
\caption{These are scatter plots of the combined and coherent SNRs of injection triggers, represented by red plus symbols, and background (or ``slide'') triggers, represented by the black crosses. The coherent SNR was used to cluster the triggers, from both injections and slides. The coherent SNR performs noticeably better than the combined effective SNR in discriminating signals from background: In the left plot, at a detection threshold of a little above 6 in the coherent SNR all the injections found in the coincident stage are recovered with a vanishing false-alarm probability. For the same false-alarm probability, the combined effective SNR detects a lesser number of injected signals. 
}
\label{fig:combinedVsCoherentSNR}
\end{figure*}

To study the performance gain arising from using the coherent stage, we ran the 
CBC search pipeline with and without that stage on simulated Gaussian
noise, with LIGO-I noise PSD \cite{lal} in the 4km LIGO detectors in Hanford (H1), 
Livingston (L1), and in the Virgo detector (V1), for the duration of 
approximately a month. 
(A similar study is being conducted for networks where the advanced-LIGO and advanced Virgo design sensitivities will be used for the LIGO and Virgo detectors, respectively, including a possible LIGO detector in Australia \cite{Ghosh}.)
Specifically, this search pipeline was run once with signal injections and again (parallelly) without injections but with time-slid data so that the background could be estimated. 
The left plot in Fig. \ref{fig:combinedVsCoherentSNR} compares the performance of the coherent statistics and the combined effective SNR. The right plot there compares the coherent SNR and null-stream statistics. For these simulations, 1051 signals were injected in software
in all three detectors. The source distances of all injections were between 100-500 Mpc. The total masses of these sources were chosen to be in the range 25-100 $M_\odot$, and component masses between 1-99 $M_\odot$. A total of 55 of those injections were found, above the single-interferometer detection thresholds of 5.0 and coherent SNR threshold of 3.75\footnote{The detection probabilities are small because, first, all injections made were weak and, second, here we focused only on triggers that are coincident in all three detectors. Owing to sensitivity disparities, it is more likely to find injection trigger coincidences in two of the three detectors \cite{Ghosh}. Only weak injections were made since that is where the coherent code can help improve the performance of current searches.}. The latter threshold was intentionally chosen to be lower since we anticipated that some coincident background triggers will have negative cross-terms owing to incoherent phases, thereby, yielding lower coherent SNRs.

\begin{figure*}[tb]
\centering
\includegraphics[width=8.5cm]{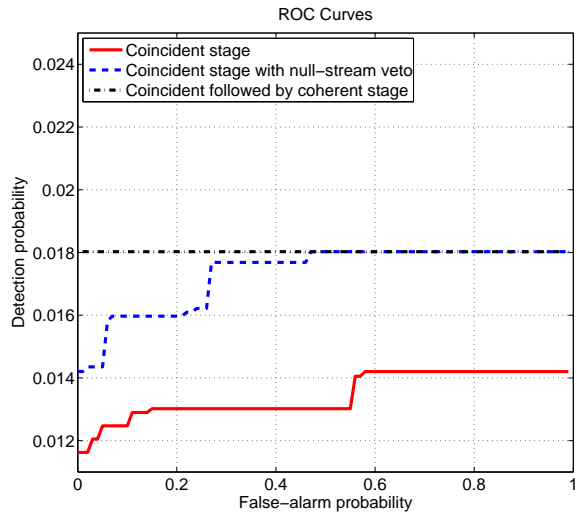}
\caption{The receiver operating characteristic (ROC) curves of three CBC searches are compared above. The ROC of the search with the coincident stage alone is plotted in solid red line, and has the weakest performance owing to the 13 found injections that are weaker than the loudest background trigger in that search. On the other hand, the ROC curve for the hierarchical pipeline, with coherent stage included, is shown in black dash-dotted line and has the best performance. It has a constant detection probability because all found injections are louder than the loudest background trigger for this pipeline. Finally, the third ROC curve, shown as a blue dashed line is the coincident stage, with the null-stream veto applied. This veto improves the performance of the coincident pipeline, so much so that for low detection-thresholds (or high false-alarm probability) its ROC curve rises to match that of the pipeline with the coherent stage. The average error in the detection probabilities plotted here is less than $3\times 10^{-4}$. 
}
\label{fig:compareRocs}
\end{figure*}

All injections recovered by the coincident stage were also found by the coherent stage, and are symbolized by red pluses. The black crosses depict the background triggers that are found by the coincident stage and survive the coherent stage. The blue circles, on the other hand, denote background triggers in the coincident stage that got vetoed by the choice of the threshold on the coherent SNR in the coherent stage. To include them in the left plot, we arbitrarily assign all of them $\rho_{\rm coh} = 3.0$. Comparing the sets of black crosses and blue circles reveals that the coherent stage not only reduces the number of background triggers but, in this case, also vetoes some of the loudest ones (in combined-effective SNR).
Furthermore, whereas all found injections have coherent SNR greater than that of the loudest background trigger, 13 of them have combined-effective-SNR weaker than that of the loudest background trigger (shown in blue circles). When compared to the loudest black cross, that number drops to 7. It drops further when some of the background triggers with the loudest null-stream (as shown in the right plot) are vetoed. The resulting performance improvement is depicted in the blue dash-dotted Receiver-Operating-Characteristic (ROC) curve in Fig. \ref{fig:compareRocs}; its performance is better than that of the coincident stage (shown in red), without the null-stream vetoes. The former asymptotes to the ROC curve of the coherent stage (shown in black dashes) for higher false-alarm probabilities.

Finally, Fig. \ref{fig:combinedVsCoherentSNR} reveals the existence of a gap between the loudest background and the weakest injection $\rho_{\rm coh}$ values. One might argue that this is owing to the lack of a sufficient number of weak signal injections made into the data. We have verified that, indeed, one can get some injection triggers to show up in that gap by making multiple weak injections (say, with source distances between 500-750 Mpc) in the data. Those studies also reveal that the detection efficiency in that region is very low (i.e., less than 1 in 250). We believe that this low efficiency is partly caused by the coincident stage, in the way it has been designed and tuned, acting as a bottleneck for the coherent stage. 


\section{Discussion}
\label{sec:discussion}

The main advantage of implementing a blind coherent search in the hierarchical manner explained above is that it has a lower computational cost compared to that of a fully coherent search pipeline. This is primarily because it reduces the number of time-of-arrival values for the coherent code to search for, and because recognizing coincidences is relatively cheaper computationally. There are additional reasons, such the inherent detector-bound nature of data-quality cuts, 
which are best implemented in the matched-filtering stage. This in turn can reduce an otherwise triple-coincident trigger into a double-coincident one if the third IFO data-points around the concurrent time get vetoed. Since the coincident and coherent statistics are the same for two-site CBC searches, it makes sense to not follow them up with the coherent stage.

There are, however, some demerits of searching hierarchically. The first one came to the fore in the results presented above, where the coincident stage is potentially affecting the efficiency of the coherent stage in finding injections. Indeed, it may be possible to improve the injection finding efficiency by reducing the SNR thresholds in the matched-filtering step of the coherent stage. While that may happen, it is also likely that the overall performance of the pipeline will be hurt since it will tend to increase the background rate as well. An alternative solution is to retain the original mass-pairs of the coincident triggers in the coherent stage instead of replacing them with {\it max}-SNR mass-pairs. This will ensure that injection-finding efficiency of the matched-filtering stage is unaffected, but may hurt the coherence of the triggers and, therefore, ultimately affect the injection finding efficiency of the coherent stage. It may also cause the false-alarm rate to rise, owing to the less stringent requirements on the agreement of the mass-pair values across the network of detectors. 

A more optimal solution that addresses the drawbacks of the last two solutions is to assign to every coincident trigger multiple mass-pair templates to search the data with in the coherent stage. This approach makes sense since statistical errors alone are known to cause substantially different mass-templates to be triggered by signals in different detectors arising from the same (injected simulated) source. However, as was shown by the work in Ref. \cite{Robinson:2008un} on identifying coincidences, the separation in the mass parameter-space between triggers in two detectors from the same source can be wide enough to allow for multiple other mass templates to fit in between. Some of these intermediate mass-templates can have a greater chance of not only passing the SNR threshold in individual detectors but also appearing as coherent. The main problem to attack here is to 
find what the optimal density and size are of these relatively small template banks localized around the coincident mass-pairs. Too small a density or size can hurt signal-finding efficiency and too big a density or size can increase the background rate. This is the subject of another study in progress \cite{Ghosh}. 

\acknowledgments 

We would like to thank the numerous members of the LIGO Scientific Collaboration and the Virgo Collaboration who have helped in testing and maintaining the coherent CBC code in LAL. 
Specifically, we thank Shawn Seader, Aaron Rogan, Duncan Brown, Jolien Creighton, Steve Fairhurst, Drew Keppel, and Eirini Messaritaki for discussions and contributions to the code in LAL. We also thank Malik Rakhmanov for carefully reading the manuscript and making helpful comments on it. One of us (SB) is grateful to Alan Weinstein for his hospitality in Caltech where some of this work was done. Three of us (SB, SG and DT) would also like to thank Bruce Allen for his hospitality during our stay at Hannover. A large fraction of the computations reported in this paper were performed on the Atlas supercomputing cluster at the Albert Einstein Institute, Hannover. We would also like to thank Stuart Anderson, Carsten Aulbert, Oliver Bock, Kipp Cannon, Collin Capano, Alex Dietz, Henning Fehrmann, Nick Fotopoulos, Romain Gouaty, Chad Hanna, Ian Harry, Sergei Klimenko, Badri Krishnan, Greg Mendell, Reinhard Prix, Fred Raab, and B. S. Sathyaprakash for helpful discussions. It is a pleasure to thank the Perimeter Institute for hosting the Numerical Relativity and Data Analysis meeting in 2010, where some of the results presented in this work were discussed.
This work is supported in part by NSF grants PHY-0758172 and PHY-0855679.

\bibliography{References}

\end{document}